\documentclass[aps,prb,twocolumn,superscriptaddress,showpacs]{revtex4-1}
\usepackage{graphics}
\usepackage{graphicx}
\usepackage[english]{babel}
\usepackage{color}
\usepackage{amsmath}
\usepackage{mathtools}

\begin{document}

\title{Bridge in micron-sized Bi$_2$Sr$_2$CaCu$_2$O$_{8+y}$
sample act as converging lens for vortices}

\author{Joaqu\'{i}n Puig}%
\affiliation{Centro At\'{o}mico Bariloche and Instituto Balseiro,
CNEA, CONICET and Universidad Nacional de Cuyo, 8400 San Carlos de
Bariloche, Argentina}

\author{N\'estor Ren\'e  Cejas Bolecek}
\affiliation{Centro At\'{o}mico Bariloche and Instituto Balseiro,
CNEA, CONICET and Universidad Nacional de Cuyo, 8400 San Carlos de
Bariloche, Argentina}

\author{Jazm\'{i}n Arag\'{o}n S\'{a}nchez}%
\affiliation{Centro At\'{o}mico Bariloche and Instituto Balseiro,
CNEA, CONICET and Universidad Nacional de Cuyo, 8400 San Carlos de
Bariloche, Argentina}

\author{Moira In\'es Dolz}
\affiliation{Departamento de F\'{i}sica, Universidad Nacional de San
Luis and CONICET,  San Luis, Argentina}

\author{Marcin Konczykowski}
\affiliation{Laboratoire des Solides Irradi\'es, Ecole
Polytechnique, CNRS, 91128 Palaiseau, France.}

\author{Yanina Fasano}
\affiliation{Centro At\'{o}mico Bariloche and Instituto Balseiro,
CNEA, CONICET and Universidad Nacional de Cuyo, 8400 San Carlos de
Bariloche, Argentina}

\date{\today}

\begin{abstract}
We report on direct imaging of vortex matter nucleated in
micron-sized Bi$_2$Sr$_2$CaCu$_2$O$_{8+y}$ superconducting samples
that incidentally present a bridge structure. We find that when
nucleating vortices in a field-cooling condition the deck of the
bridge acts as a converging lens for vortices. By means of Bitter
decoration images allowing us to quantify the enhancement of
vortex-vortex interaction energy per unit length in the deck of the
bridge, we are able to estimate that the deck is thinner than $\sim
0.6 $\,$\mu$m. We show that the structural properties of vortex
matter nucleated in micron-sized thin samples are not significantly
affected by sample-thickness variations of the order of half a
micron, an important information for  type-II superconductors-based
mesoscopic technological devices.

\end{abstract}

 \keywords{}
 \maketitle

\section{Introduction}

The miniaturization of technological devices have driven the study
on how the confinement and surface effects alter the physical
properties of systems of interacting
objects.~\cite{Coombes1972,Goldstein1992,Tolbert1994,Guisbiers2009,Dolz2019}
Nucleating small crystals of vortex matter in micron-sized
superconducting samples gives us the possibility of studying this
general problem in a playground where the relevant parameters are
easily controlled. Indeed, the density of interacting objects can be
tuned by applied field, the interaction of vortices with the
underlying disorder of the host sample can be altered by changing
temperature, and the confinement and surface effects can be tailored
by a suitable design of the
samples.~\cite{Moshchalkov1995,Geim1997,Palacios1998,Schweigert1998,Wang2001,Dolz2015}
There is sufficient evidence in the literature that size and surface
effects in micron-sized superconducting samples produce dramatic
changes on their physical properties. For instance, for type-I
superconductors of one-micron or sub-micron size, the boundary
conditions imposed by the shape and geometry of the sample to the
quantization of angular momentum of the Cooper pair wave function
govern the superconducting phase boundary, and stabilizes a complex
phase diagram with several confinement-induced
transitions.~\cite{Moshchalkov1995,Geim1997} In the case of
micron-sized type-II superconductors, changing the applied field
induces not only different regimes in the magnetic response of the
system, but can also trigger a structural transition in vortex
matter from a giant vortex state with multiple flux quanta to a
glass state with single-fluxoid
vortices.~\cite{Palacios1998,Schweigert1998}

On increasing the sample size to tens of microns, and then
nucleating crystals with few hundred vortices, some works report on
how the thermodynamic, magnetic, and structural properties of vortex
matter are significantly different from the bulk
case.~\cite{Dolz2019,Wang2001,Dolz2015,Wang2002,Dolz2014b,Cejas2015,Ooi2017,Cejas2017}
These works warn on a need of better characterizing these properties
for potential applications of these systems to miniaturized
superconducting devices. Regarding the structural properties, a
proliferation of topological defects induced by confinement has been
reported for crystals with few hundred vortices in the case of the
layered Bi$_2$Sr$_2$CaCu$_2$O$_{8+y}$
system.~\cite{Cejas2015,Cejas2017} For thin micron-sized disks,
topological defects in the vortex structure proliferate on
decreasing the radius from 50 to 30\,$\mu$m and also present an
inhomogeneous spatial distribution.~\cite{Cejas2017} At the vicinity
of the edge, the number of defects amplifies over a characteristic
length in which vortex rows tend to bend mimicking the shape of the
sample. In contrast, within the center of the samples the positional
order of the vortex structure is consistent with the Bragg-glass
phase. This healing length at which topological defects are cured
towards the center may be a key quantity to model confinement
effects in vortex matter nucleated in micron-sized samples. The
geometry of samples is also another parameter that affects the
structural properties of vortex crystals at the mesoscopic scale.
For example, a smaller density of topological defects is nucleated
in square than disk thin mesoscopic samples of the same typical
size. In addition, in the former case the vortex rows accommodate
parallel to the edges without bending.~\cite{Cejas2015}

Previous studies have reported data in samples with a constant
thickness,  then very little is known about the effect of varying
the sample thickness in micron-sized samples. In order to amplify
the effect of sample thickness variations, it is desirable to choose
a superconducting material with large line energy per unit length,
$\varepsilon_{L}=(\Phi_{0}/4 \pi \lambda)^{2} \ln{\kappa}$. A
material such as Bi$_2$Sr$_2$CaCu$_2$O$_{8+y}$, with $\kappa \approx
200$, seems a suitable candidate for studying this issue. Vortex
matter in bulk samples of this material presents a rich phase
diagram that includes liquid and glassy phases with different
magnetic~\cite{Pastoriza,Zeldov,Nelson,Avraham,Dolz2014} and
structural
properties~\cite{Fasano1999,Fasano2005,AragonSanchez2019,AragonSanchez2020}
depending on the type of disorder of the host sample. In this paper
we study the structural properties of vortex matter in micron-sized
thickness-modulated samples of the extremely-anisotropic pristine
Bi$_2$Sr$_2$CaCu$_2$O$_{8+y}$ superconductor. We directly image the
$\sim 700$-vortices crystal nucleated in a square thin sample of
roughly 40\,$\mu$m side and 2\,$\mu$m thickness that incidentally
presents a bridge structure. We find that when nucleating vortices
in a field-cooling condition the deck of the bridge acts as a
converging lens for vortices.  Irrespective of the higher vortex
density in the bridge than in the base, the structural properties of
vortex matter in both regions are not significantly dissimilar. By
means of energetic arguments we estimate that the deck of the bridge
is $\sim 0.6\,\mu$m thick or even thinner. This quantitative
estimation allows us to measure a top limit on how thin can be the
samples in order for the structure of vortex crystals in mesoscopic
thin samples not being significantly affected by sample-thickness
variations and still behaving as a three-dimensional structure.

\section{Method}

\begin{figure}[bbb]
    \centering
    \includegraphics[width=\linewidth]{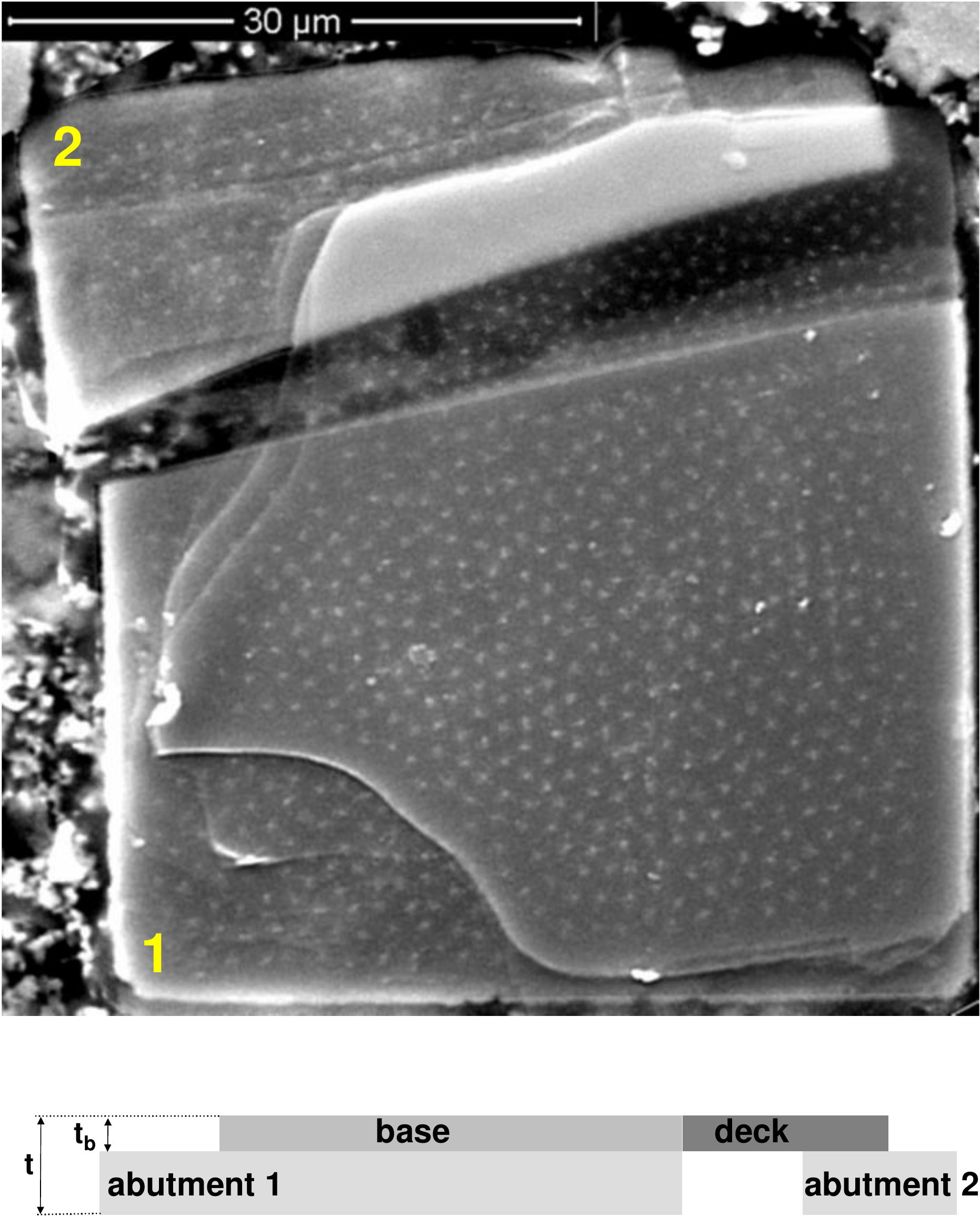}
    \caption{Scanning electron microscope image of the field-cooling magnetic
    decoration performed at 12\,Oe and 4.2\,K in a micron-sized
    square thin sample of Bi$_2$Sr$_2$CaCu$_2$O$_{8+y}$
    presenting a bridge structure. Fe clusters (white dots) decorate vortices
    as they impinge at the sample surface. The sample has a nominal size of
    $\sim 40$\,$\mu$m side and $2\,\mu$m thickness $t$. Bottom: schematic lateral
    representation of the bridge structure  indicating the parts of the sample that
     act as  deck, base, and abutments of the bridge. The abutments 1 and 2 are also
     labeled with yellow numbers in the top picture. The thickness of the deck
     is indicated as $t_{b}$.}
    \label{figure1}
\end{figure}

The micron-sized square thin sample with a bridge structure was
engineered from a bulk,  pristine, and optimally-doped
 Bi$_2$Sr$_2$CaCu$_2$O$_{8+y}$ crystal with a critical temperature $T_{\rm
 c}=90$\,K. The sample was obtained following a top-down procedure that
 combines optical lithography with physical ion-milling,
 see Refs.\,\onlinecite{Dolz2015,Dolz2019}
 for further technical details. With the aim of directly imaging the vortex
 structure nucleated in such samples, we cleave the
 physically-etched bulk crystal, then manipulate one by one the micron-sized samples
 and glue them with conducting epoxy onto a scanning electron microscope sample holder.
 The sample studied in this work belongs to a batch of square thin samples with
 nominally 40\,$\mu$m side and $t=2\,\mu$m thickness.
During the gluing process, pressure is applied with a
micro-manipulator in order to ensure good thermal and electrical
conductivity. Since Bi$_2$Sr$_2$CaCu$_2$O$_{8+y}$ is an easily
cleavable and fragile material, many micron-sized samples break or
crack during the gluing process. By serendipity, the one that we
study in this paper broke, but only in the bottom part. During the
gluing process both pieces, labeled as abutments 1 and 2 in
Fig.\,\ref{figure1}, were in-plane split less than 10\,$\mu$m. The
top part of the sample remained attached to the larger abutment,
number 1. As shown in Fig.\,\ref{figure1}, this resulted in a sample
with a bridge-like structure   with a trapezoidal bridge deck of
$\sim 40$\,$\mu$m width and $5-10$\,$\mu$m length. The base of the
bridge is the larger region of the sample observed at the
middle-bottom of Fig.\,\ref{figure1} and presents an irregular and
curved step (see bottom-left) that exposes the abutment 1 to which
the base remained attached. The deck of the bridge has a thickness
$t_{\rm b}$ and in the base region the sample has the nominal
thickness $t \sim 2$\,$\mu$m, see the schematic lateral
representation of the sample at the bottom of Fig.\,\ref{figure1}.
The part of the sample where the deck settles on top of abutment 2
is observed as a very bright region in this scanning electron
microscopy (SEM) image. The difference in contrast with the rest of
the sample might come from this region being charged by the electron
beam, namely not being able to discharge the electrons to the sample
holder (ground) due to a bad electrical conductivity with the rest
of the sample. It is rather ubiquitous to observe this flake-like
edges when cleaving  layered Bi$_2$Sr$_2$CaCu$_2$O$_{8+y}$ samples.

We image the vortex structure in real space and with single vortex
resolution by means of the magnetic decoration
technique.~\cite{Fasano2003} For the case studied here, we nucleated
the vortex structure in a field-cooling condition applying a field
of 12\,Oe  in the normal state and cooling down to the decoration
temperature of 4.2\,K. The vortex decoration consists in evaporating
$\sim 100$\,nm size Fe particles in a sealed chamber submerged in a
$^{4}$He bath. Inside the chamber the Fe particles collide and form
clusters of size controlled by the He-exchange gas pressure. The Fe
nanoparticles are attracted towards the vortex cores impinging at
the sample surface due to the magnetic force generated by the
vortex-induced local field gradient. Once the Fe particles decorate
the vortex positions, the sample is warmed up and the Fe particles
that remain attached to the surface due to van der Waals forces (see
white dots in Fig.\,\ref{figure1}) are observed with a scanning
electron microscope at room temperature.

\section{Results}

\begin{figure}[bbb]
    \centering
    \includegraphics[width=\linewidth]{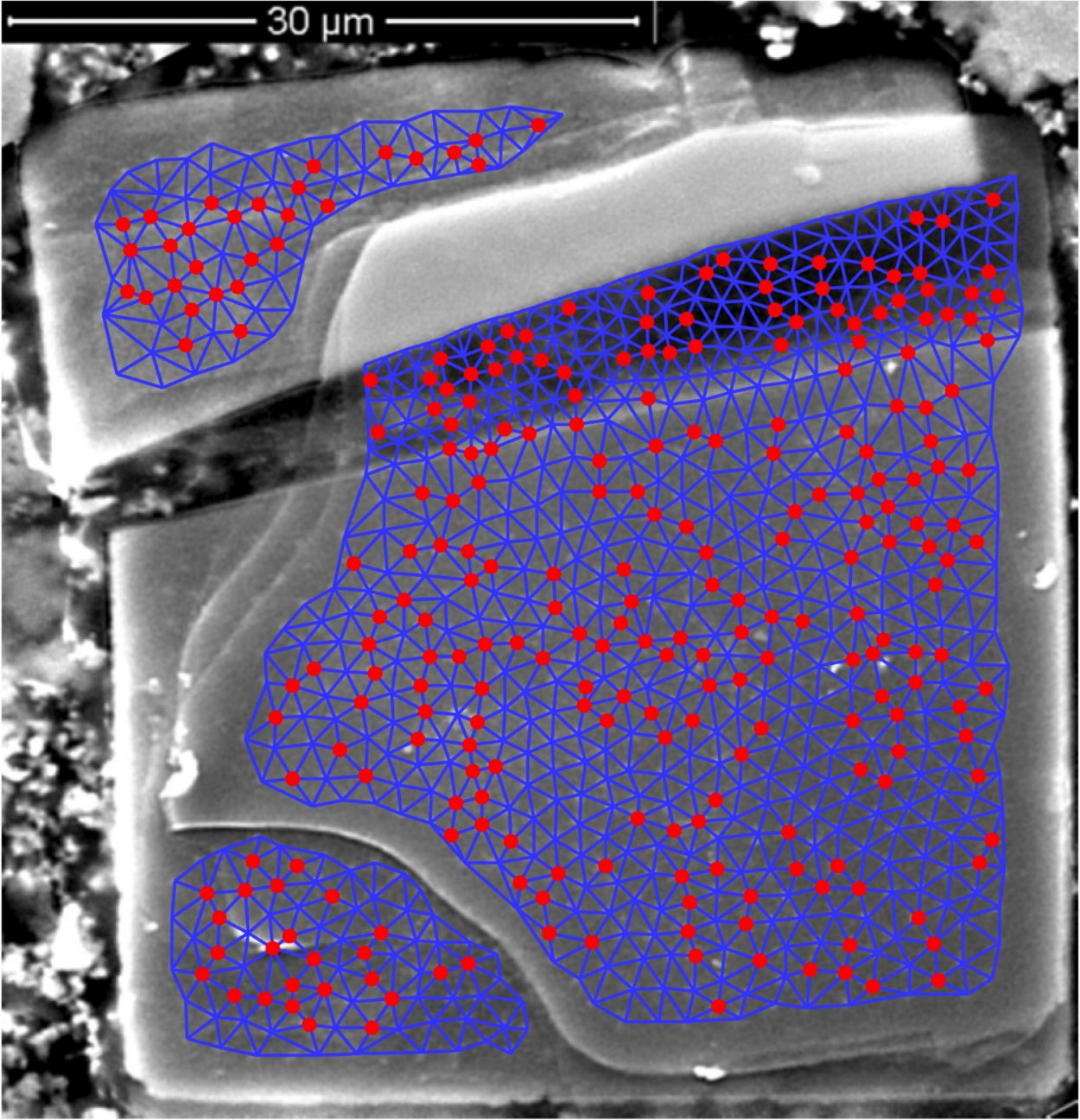}
    \caption{Delaunay triangulation of the vortex structure nucleated in the micron-sized
    square sample of Bi$_2$Sr$_2$CaCu$_2$O$_{8+y}$
    presenting a bridge structure. First-neighbor vortices are bond with blue
    lines and non-sixfold coordinated vortices generating topological
    defects are highlighted in red.}
    \label{figure2}
\end{figure}

\begin{figure}[bbb]
    \centering
    \includegraphics[width=\linewidth]{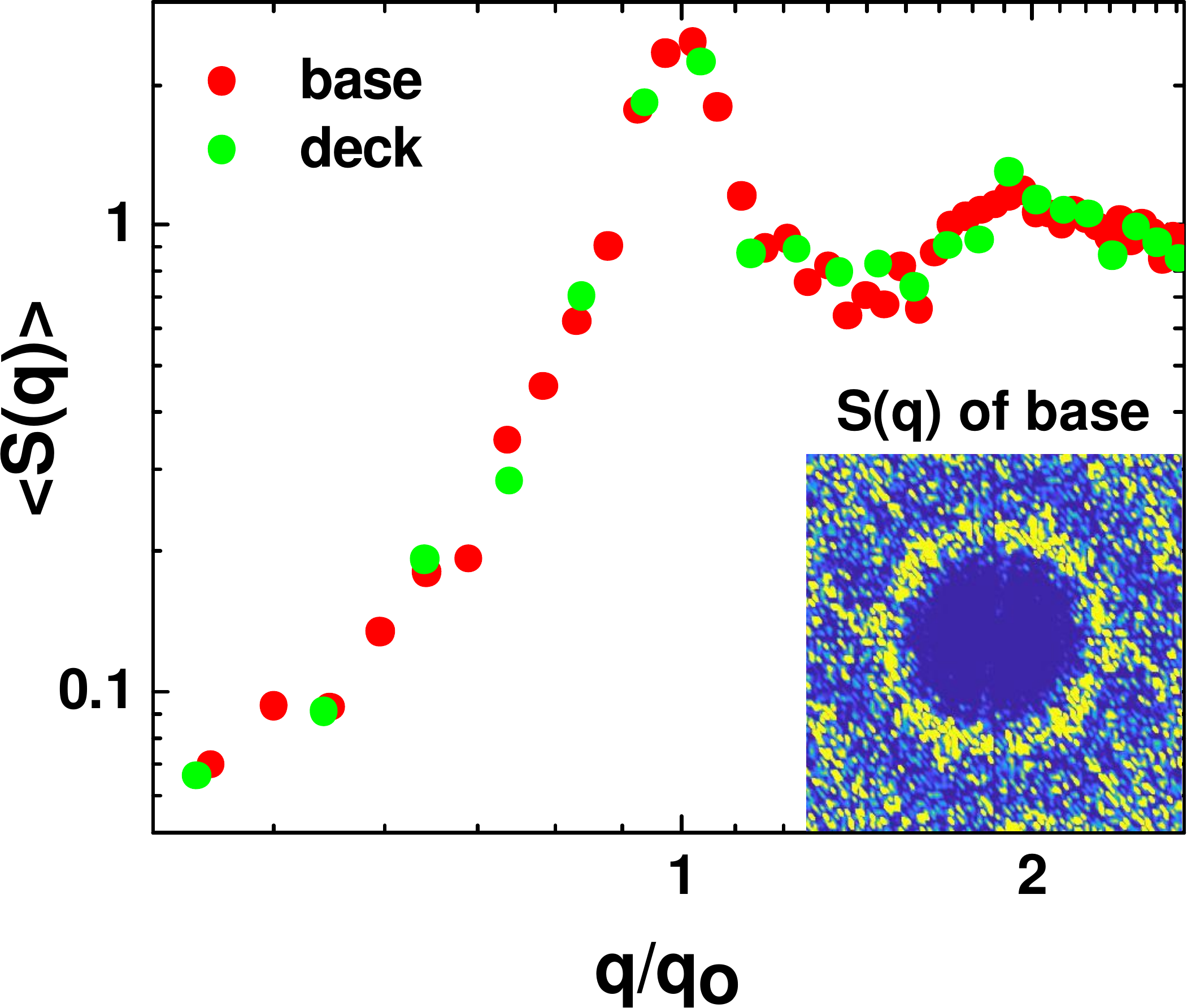}
    \caption{Main panel: Angularly-averaged structure factor of the
    vortex structure nucleated in the base (red points) and deck (green points)
     regions of the micron-sized square sample of Bi$_2$Sr$_2$CaCu$_2$O$_{8+y}$
    presenting a bridge structure. These data are obtained from
    angularly-averaging the two-dimensional structure factor $S(q)$ at a
    given $q \pm \delta q$ circular region. Insert: Example of the $S(q)$
    for vortices nucleated in the base region of the sample.}
    \label{figure3}
\end{figure}

Figure \ref{figure1} shows a snapshot of the vortex structure
nucleated in the micron-sized Bi$_2$Sr$_2$CaCu$_2$O$_{8+y}$ square
sample with a bridge structure, following a field-cooling process at
an applied field of 12\,Oe. The picture is obtained from a magnetic
decoration performed at 4.2\,K and vortices are observed as white
dots in the SEM image of Fig.\,\ref{figure1}. The most notorious
fact of this snapshot is that the bridge structure acts as a
converging lens for the vortex structure: The average vortex density
in the deck is 15\,G whereas in the base is
 10\,G, and in abutments 1 and 2 is of 10 and 9\,G, respectively.
 The average vortex density in the whole sample  is 11\,G.
 The region where the deck settles on
 top of abutment 2  is observed as
 a bright region where vortices (white dots) are not resolved
from  the background. Figure\,\ref{figure2} shows that only the
vortices located at the very bottom part of this region are
resolved. This might be due to the sample getting charged by the
electron beam due to poor electrical contact with the below
abutment, as mentioned.

Figure\,\ref{figure2} shows the same image than in
Fig.\,\ref{figure1} but  with the Delaunay triangulation of the
vortex structure superimposed. The triangulation shows each vortex
bonded to its first-neighbors with blue lines. Non-sixfold
coordinated vortices are highlighted with red circles. This figure
makes clear that at the bottom of the deck of the  bridge vortices
are aligned along it. Nevertheless, this construction also shows
that there is not a strong correlation between the nucleation of
grain boundaries and variations in the sample thickness, either due
to steps as in the bottom-left part, or to the thickness reduction
at the deck of the bridge. The vortex structure is polycrystalline
in the whole sample with crystallites formed by 10-20 vortices. The
density of topological defects, namely the ratio of the non-sixfold
coordinated to the total number of vortices, is of $46$\,\% in the
deck and $34$\,\% in the base regions of the sample. Most of
topological defects form grain boundaries and a few isolated edge
dislocations. Then, the focusing effect produced in the deck of the
bridge entails a moderate enhancement of topological defects in the
deck, probably produced by the combined effect of the density change
with a structure aligned along the bottom part of the deck.

It is important to point out that the vortex structure nucleated at
12\,Oe in bulk Bi$_2$Sr$_2$CaCu$_2$O$_{8+y}$ samples is also
polycrystalline since vortex interaction is weak at such low
fields.~\cite{AragonSanchez2019,AragonSanchez2020} Nevertheless, for
a vortex structure with larger interaction energy (applied fields of
$\sim 40$\,Oe), previous works in bulk Bi$_2$Sr$_2$CaCu$_2$O$_{8+y}$
samples show that steps in the surface act as grain-boundary
nucleation centers,~\cite{Dai1994} in contrast to data in
NbSe$_2$.~\cite{Pardo1997} This difference in vortex structure
behavior in both materials has been ascribed to the larger line
energy (larger $\kappa$) of Bi$_2$Sr$_2$CaCu$_2$O$_{8+y}$ than of
NbSe$_2$.~\cite{Pardo1997}

Aside from the converging lens effect of the bridge structure and
the difference in topological defects, the positional order of
vortex matter nucleated in the deck and the base with different
thicknesses are alike. We characterize the positional order
computing the angularly-averaged structure factor $\langle
S(q)\rangle$ of vortex matter in both regions. This magnitude
results from angularly-averaging the two-dimensional structure
factor $S(q)$ at a given $q \pm \delta q$ circular region. The
latter is obtained by computing $S(q)= \mid
\rho(q_{x},q_{y})\mid^{2}_{z=0}$, with $ \rho(q_{x},q_{y})$ the
Fourier transform of the discrete vortex density detected at the
sample surface. An example of $S(q)$ is shown in the insert to
Fig.\,\ref{figure3}. The angularly-averaged structure factor for the
deck and base regions are alike: Both present a peak at the average
first-neighbor distance $a$ (traduced in reciprocal space as
$q/q_{0}=1$ with $q=2\pi/r$ and $q_{0}=2\pi/a$ the Bragg
wave-vector), and decay algebraically at low $q$-values. These
findings are presented in the main panel of Fig.\,\ref{figure3}
showing the $\langle S(q)\rangle$ for both regions of the sample.

\section{Discussion}

 The focusing effect occurs probably at the beginning of the field
cooling process. Even though the decoration of vortex positions is
performed at 4.2\,K, in a field-cooling process this technique
actually captures a snapshot of the vortex structure frozen at a
temperature $T_{\rm freez}$ close to the irreversibility
line.~\cite{Fasano2005} At the irreversibility temperature $T_{\rm
irr}$, weak bulk pinning sets in and then the vortex structure is
frozen at lengthscales of the lattice spacing $a$. On further
cooling, vortices are able to explore the sample only within
distances smaller than the range of the pinning potential, $\xi \ll
\lambda$, with $\lambda$ the penetration depth and also the spatial
resolution of the magnetic decoration technique. On nucleating
vortices at high temperatures, in the liquid vortex phase bulk
pinning is not relevant and the nucleation of a denser structure in
the deck can be understood just by considering vortex-vortex
interaction energy arguments exclusively. Since the  interaction
energy between vortices is equal to the interaction energy per unit
length, $E_{\rm int}$, times the thickness of the region of the
sample, a denser structure with larger $E_{\rm int}$ can be
stabilized in a significantly thinner region of the sample.

The possibility of this focusing effect being marginally produced by
a stronger screening in the thicker abutments has to be considered.
A sample-thickness variation induced screening effect was long ago
observed in field-ramping experiments in superconducting weak links
of Al thin films.~\cite{Dolan1977} Indeed, the flux penetration can
be inhomogeneous in thin mesoscopic samples of thickness $t$ and
width $2w$ comparable to $\lambda$ since the London limit with
$\lambda \rightarrow 0$ is no longer applicable for these
geometries.~\cite{Landau1960} In these type of samples, within the
London model but taking $\lambda >0$, the relevant length-scale for
magnetic field penetration is the Pearl length
$\Lambda=\lambda^{2}/t$. For some particular sample geometries and
$\Lambda/w$ ratios, numerical solutions to the thickness-dependent
current distribution in thin mesoscopic samples  have been
theoretically
obtained.~\cite{Brandt2001,Plourde2001,Mawatari2012,Via2013} These
works consider the case of flux penetration in the Meissner
state~\cite{Mawatari2012,Via2013}, or for low fields close to the
Meissner-mixed state transition~\cite{Brandt2001}, and all of them
study flux penetration in the non-static case of ramping
field.~\cite{Brandt2001,Plourde2001,Mawatari2012,Via2013} In
contrast, the experimental situation reported here correspond to
vortices nucleated at fields way larger than the effective $H_{\rm
c1}$ at high temperatures, then cooling without changing the vortex
density. For the presently studied field-cooling experimental
situation, the effect of a thickness-variation-induced inhomogeneous
flux penetration due to an enhancement of screening currents in the
thicker region of the sample is expected to be significantly
reduced. Detailed theoretical work in the mixed state in
field-cooling conditions is mandatory to asses if the mentioned
effect is suppressed completely, but since that is beyond the aim of
our paper we will consider as expected that it is a marginal effect.

\begin{figure*}[ttt]
\centering
\includegraphics[width=2\columnwidth]{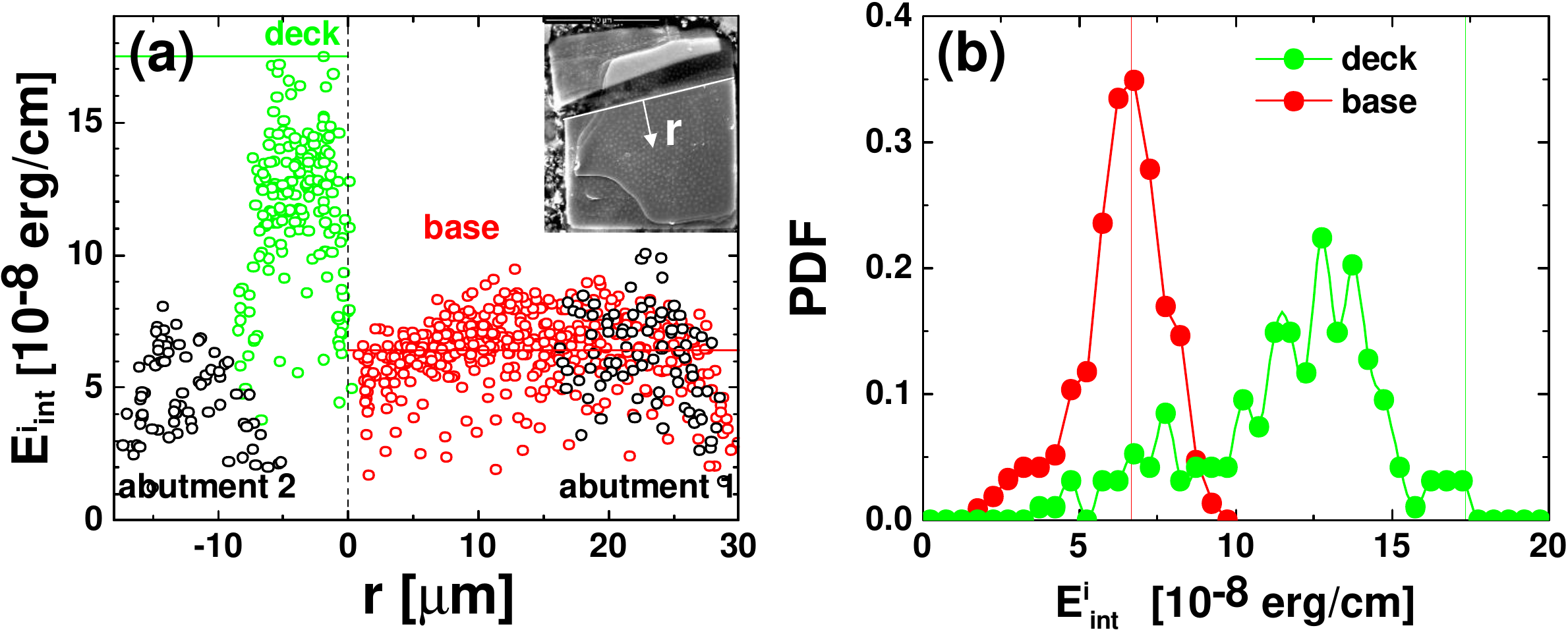}
    \caption{Vortex-vortex interaction energy of the
    structure nucleated in the micron-sized square thin
    Bi$_2$Sr$_2$CaCu$_2$O$_{8+y}$ sample
    presenting a bridge structure. (a) Main panel: Interaction energy per unit length
    of individual vortices shown as a function of the distance $r$ between each
    vortex and
    the edge of the deck (see insert). Data result from computing the
    interaction of every vortex with the rest of vortices nucleated
    in the sample (see text for further details). Data in different colors correspond
    to different regions of the sample as indicated with the legends. Lines are the
    average value of the interaction energy for the
    base (red) and the maximum for the deck (green) regions. (b) Probability density
    function of the
    vortex-vortex interaction energy per unit length in the deck (green points) and base (red points)
    regions. }
    \label{figure4}
\end{figure*}

The focusing effect entails a local decrease of inter-vortex
distance and then an enhancement of vortex-vortex interaction energy
per unit length, $E_{\rm int}$. For every single i-th vortex, in the
case of a large $\kappa = \lambda/\xi$ superconductor such as
Bi$_2$Sr$_2$CaCu$_2$O$_{8+y}$, and for low vortex densities as in
our case, this magnitude can be computed as

$$
E^{i}_{\rm int} (T_{\rm freez})= 2 \epsilon_{\rm 0} \sum_{j} K_{0}
(\frac{\mid r_{ij} \mid}{\lambda(T_{\rm freez})}).
$$

\noindent In this expression $\epsilon_{\rm 0}= \Phi_{0}/(4 \pi
\lambda(T_{\rm freez}))^{2}$ is the vortex line tension, $\Phi_{0}=
2.07\cdot 10^{-7}$\,G$\cdot$cm$^2$ the magnetic flux quantum,
$K_{0}(x)$ is the lowest-order modified Bessel function, and $\mid
r_{ij} \mid$ the separation between vortex $i$ and its neighbors
$j$. The sum in this expression runs over all the $N-1$ remaining
vortices of the structure. In macroscopic samples this sum becomes
numerically expensive to compute and since $K_{0}(x)$ strongly
decays with $x$, typically a proper cut-off is considered in the
calculation. In the case of our structure of $\sim 700$ vortices, we
considered all the digitalized vortices shown in Fig.\,\ref{figure2}
to calculate $E^{i}_{\rm int}$. For this calculation we have
considered $\lambda(T_{\rm freez})=0.6$\,$\mu$m. This value is
obtained from taking into account the $\lambda(T/T_{\rm c})$
evolution reported in Ref.\,\onlinecite{Kees} for bulk pristine
Bi$_2$Sr$_2$CaCu$_2$O$_{8+y}$ samples, and from considering that
$T_{\rm freez} \sim T_{\rm irr} = 89$\,K for a vortex structure
nucleated with an applied field of 12\,Oe.~\cite{Dolz2014}

 Figure\,\ref{figure4} (a)
shows the interaction energy per unit length of every vortex sorted
as a function of the distance $r$ between the vortex and the bottom
part of the deck ($r=0$). The insert to this figure shows a
schematic representation on how this distance is computed. As
indicated with the legends, red data points correspond to vortices
in the base region, green ones to vortices in the deck and black
ones to the abutments 1 and 2. Another token of the focusing effect
in the deck of the bridge is that $E^{i}_{\rm int}$ is significantly
larger for vortices there than in the base and abutments regions.
Indeed, the average value of this magnitude is 12.2 in the deck
against 6.4\,$\times 10^{-8}$\,erg/cm in the base (see red line). A
similar $E_{\rm int}$ mean value of 6.4, and a slightly smaller of
$\sim 5$\,$\times 10^{-8}$\,erg/cm, are found in the abutments 1 and
2 regions, respectively. This implies that the thicknesses of the
abutments are quite close to that of the base region.  The panel (b)
of Fig.\,\ref{figure4} shows the probability density function (PDF)
of the distribution of $E^{i}_{\rm int}$ for individual vortices in
the deck and base regions (green and red points respectively). As
already visible in panel (a), the distribution of this magnitude has
not only a larger mean value but also a larger standard deviation in
the deck (24\,\%) than in the base (18\,\%).

Since at the freezing temperature of the vortex structure the volume
interaction energy has to be roughly the same in the whole sample,
the difference in the interaction energy per unit length in the deck
and base regions can be taken into account to estimate the thickness
of the deck  $t_{\rm b}$. Then, the maximum energy observed in the
deck times its thickness can not be larger than the volume energy in
the base, namely

$$
Max[E^{d}_{\rm int}] \cdot t_{\rm b} \lesssim  \langle E^{b}_{\rm
int} \rangle \cdot  t
$$

\noindent where we have estimated the volume energy in the base from
the average value of the interaction energy per unit length,
$\langle E^{b}_{\rm int} \rangle$.   Taking into account the maximum
and average values shown with green and red lines in
Fig.\,\ref{figure4}, we get $t_{\rm b} \lesssim 0.3 t \sim 0.6
$\,$\mu$m considering the nominal thickness of the sample. It is
important to realize that this numerical value for the upper bound
of $t_{\rm b}$ is in addition overestimated: Due to the lack in
contrast in the SEM picture we were not able to digitalize all
vortices in the brighter region on top of the deck, and then the
$E^{i}_{\rm int}$ for vortices in the deck is underestimated since
the contribution of half of its neighbors is missing. Nevertheless,
since $K_{0}(x)$ rapidly decays at even two lattice spacings, the
contribution from the non-detected vortices would imply only a $\sim
10-30$\,\% enhancement of interaction energy for vortices in the
deck. We estimated this percentage by simulating different
configurations of artificially-added vortices in the brighter region
with a density equal to that of the base region. We would like to
point out that similar energetic arguments have been considered by
other authors in the case of blind microholes~\cite{Bezryadin1996}
and pillar-like antipinning centers~\cite{Berdiyorov2008} in Nb
samples, but in order to explain the observed vortex arrangements,
not to quantify the sample thickness variation as we do here.

Finally, we would like to mention that  this estimation neglects any
possible effect of vortices bending along the sample, namely they
are considered as straight lines. This seems to be a reasonable
assumption since small-angle neutron scattering data in bulk
Bi$_2$Sr$_2$CaCu$_2$O$_{8+y}$ samples indicate that at low fields
the correlation length longitudinal to the direction of vortices  is
$\sim 7 a \sim 7\,\mu$m.~\cite{AragonSanchez2019} In addition, this
calculation does also not consider any difference in the vortex
pinning energy that might have the deck and base regions.

\section{Conclusion}

In conclusion, we have found that micron-sized samples of type-II
superconductors presenting a bridge-like structure might be
effective converging lens devices focusing vortex density in the
deck of the sample. Here we reveal this effect in the case of a
$\sim 700$-vortices structure with a 11\,G density nucleated in a
mesoscopic thin Bi$_2$Sr$_2$CaCu$_2$O$_{8+y}$ sample with 40\,$\mu$m
side, 2\,$\mu$m nominal thickness, and presenting a bridge with a
$\lesssim 0.6$\,$\mu$m-thick deck. We find that in this material
with large line energy this converging effect is due to the
thickness reduction at the deck allowing a larger interaction energy
per unit length while conserving the volume interaction energy.
 We also find that both, in the base and deck regions of
the sample, the structural properties of vortex matter present
similar orientational (polycrystalline) and positional order. By
means of energetic arguments we provide an (over)estimation of the
upper bound of the thickness of the deck of the bridge of $\sim 0.6
$\,$\mu$m. This thickness is also an upper bound for the structure
of mesoscopic vortex structures still behaving as three-dimensional
and not being dramatically affected by sample-thickness variations.
This is an important finding in the vortex physics of micron-sized
type-II superconducting devices intended for technological
applications: Different ranges of sample thickness and lateral size,
applied field, material anisotropy, and temperature, can produce
extremely-anisotropic vortices to behave as three or two dimensional
entities, the latter being more easy to manipulate.

\begin{acknowledgments}

This work was supported by  the Argentinean National Science
Foundation (ANPCyT) under Grant PICT 2017-2182; by the Universidad
Nacional de Cuyo research grant 06/C566-2019; and by Graduate
Research fellowships from IB-CNEA for J. P. and from CONICET for
J.P., J. A. S., and N. R. C. B. We thank to I. Artola-Vinciguerra
for assistance with SEM images, M. Li for growing the bulk single
crystals from which the samples where generated, and to H. Pastoriza
for support in the micro-engineering facilities at the CAB clean
room.
\end{acknowledgments}

\end{document}